\documentclass[twocolumn,accepted=2017-08-09]{quantumarticle}
\pdfoutput=1
\usepackage[utf8]{inputenc}
\usepackage[T1]{fontenc}
\usepackage{amsfonts,amssymb,amsmath}            
\usepackage[]{graphics,graphicx}            
\usepackage{amsthm}
\usepackage{enumerate,mathtools,bbold}
\usepackage[numbers,sort&compress]{natbib}
\usepackage{jabbrv}
    \usepackage{hyperref}

\newcommand{\ket}[1]{\left | #1 \right\rangle}
\newcommand{\bra}[1]{\left \langle #1 \right |}
\newcommand{\half}{\mbox{$\textstyle \frac{1}{2}$}}

\newcommand{\braket}[2]{\left\langle #1|#2\right\rangle}
\newcommand{\proj}[1]{\ket{#1}\bra{#1}}
\newcommand{\identity}{\mathbb{1}}
\renewcommand{\epsilon}{\varepsilon}

\begin{document}
\title{Tailoring Spin Chain Dynamics\\ for Fractional Revivals}
\author{Alastair Kay}
\date{\today}
\affiliation{Department of Mathematics, Royal Holloway University of London, Egham, Surrey, TW20 0EX, UK\\\href{mailto:alastair.kay@rhul.ac.uk}{alastair.kay@rhul.ac.uk}}
 \begin{abstract}
 The production of quantum states required for use in quantum protocols \& technologies is studied by developing the tools to re-engineer a perfect state transfer spin chain so that a separable input excitation is output over multiple sites. We concentrate in particular on cases where the excitation is superposed over a small subset of the qubits on the spin chain, known as fractional revivals, demonstrating that spin chains are capable of producing a far greater range of fractional revivals than previously known, at high speed. We also provide a numerical technique for generating chains that produce arbitrary single-excitation states, such as the $W$ state.
 \end{abstract}

\maketitle

\section{Introduction}

The task of quantum state synthesis lies at the heart of quantum technologies -- before any quantum protocol can be run, be it a Bell test \cite{clauser1969}, quantum key distribution \cite{ekert1991}, quantum cloning \cite{werner1998,kay2009,buzek1996}, random number generation \cite{pironio2010} or quantum computation \cite{raussendorf2001}, a non-trivial quantum resource, such as a Bell state, $W$-state or GHZ state must be prepared. Since the availability of this resource gives the protocol its power, it is crucial to understand how these states may best be prepared, taking into account locality constraints, control constraints etc.\ that are imposed upon a particular experiment.

To that end, we embrace the perspective of perfect state transfer \cite{bose2003,christandl2004,burgarth2005,christandl2005,kay2010a}, wherein one engineers a simple, one-dimensional system so that it accomplishes a particular task without any further user interaction. The control required of the system is restricted to the manufacturing stage, which can be verified before use. These schemes had the unexpected benefit of being up to twice as fast as the equivalent consecutive sequences of {\sc swap} gates specified by the gate model \cite{yung2006}. Once this limiting case of state transfer was established \cite{christandl2004,christandl2005,kay2010a}, a multitude of different schemes, specialised to different experimental constraints have been derived \cite{karbach2005,haselgrove2005,burgarth2005,wojcik2005}. We aim to enable this diversification for the state synthesis task. The solutions for perfect state transfer already provide examples of state synthesis by generating entanglement, both bipartite \cite{christandl2005} and that required for cluster states \cite{clark2005}, while a beautiful transformation \cite{kay2010a} of these coupling schemes permits superposition of the input state over the two extremal sites of the chain \cite{dai2010,kay2010a,banchi2015,genest2016}.

Here, we take the existing constructions for perfect state transfer and re-engineer them to produce arbitrary (one-excitation) quantum states, concentrating on the particular case of so-called fractional revivals wherein the amplitude of the final state is spread over a small number of sites on the chain. These admit the possibility of analysis (Sections \ref{sec:analysis1} and \ref{sec:fractional}), while we also provide a widely applicable numerical scheme (Section \ref{sec:numerics}), permitting the creation of $W$-states and similar, along with a starting point that appears to work well for systems of up to about 50 qubits. This complements our recent results \cite{kay2017} which showed that almost any one excitation quantum state can be created by these spin chains, with the fractional revivals being the particularly challenging cases. Moreover, in Section \ref{sec:speed} we will show that our constructions are near-optimal, achieving the desired evolution in approximately half the time required by the solutions in \cite{kay2017}, and are quite robust against imperfections (Section \ref{sec:robustness}).

\subsection{Setting}

Consider a system of size $N$, with states $\ket{1},\ldots,\ket{N}$, and a system Hamiltonian
$$
H=\sum_{n=1}^NB_n\proj{n}+\sum_{n=1}^{N-1}J_n(\ket{n}\bra{n+1}+\ket{n+1}\bra{n}).
$$
This corresponds, for example, to $N$ qubits in a line, coupled by a nearest-neighbour XX or Heisenberg Hamiltonian, restricted to the one-excitation subspace
$$
\ket{n}:=\ket{0}^{\otimes (n-1)}\ket{1}\ket{0}^{\otimes (N-n)},
$$
although there are various other mappings \cite{kay2007}, including free-fermion models such as the transverse Ising model. We denote the spectrum of $H$ by $\{\lambda_n\}$, and the corresponding eigenvectors $\ket{\lambda_n}$ have elements $\lambda_{n,1}=\braket{1}{\lambda_n}$.

Our aim is to specify the magnetic fields $\{B_n\}$ and coupling strengths $\{J_n\}$ such that the transformation
\begin{equation}
\ket{1}\xrightarrow{e^{-iHt_0}}\ket{\psi_T}=\sum_{n=1}^N\alpha_n\ket{n}\label{eqn:synthesis}
\end{equation}
is realised in a time $t_0$, where the $\alpha_n$ are all assumed to be real.

More precisely, we require that there exists some global phase $\phi$ such that
$$
e^{-iHt_0}\ket{1}=e^{i\phi}\ket{\psi_T}.
$$
Following \cite{kay2010a}, we take the inner product with an eigenvector $\ket{\lambda_n}$, giving $\bra{\lambda_n}e^{-iHt_0}\ket{1}=e^{i\phi}\braket{\lambda_n}{\psi_T}$. In other words,
$$
\lambda_{n,1}=e^{i\phi+i\lambda_nt_0}\braket{\lambda_n}{\psi_T}
$$
for all $n$. By imposing that the $\alpha_n$ are real, this can only be true if $e^{i\phi+i\lambda_nt_0}=\pm 1$ and $\lambda_{n,1}=\pm\braket{\lambda_n}{\psi_T}$, where the two equations choose the same $\pm1$ factor for each $n$. These are necessary conditions for the state synthesis task.

As perfect state transfer is a special case of state synthesis, with $\ket{\psi_T}=\ket{N}$, it is clear that these conditions are not always sufficient -- in that case, it is required that $\lambda_{n,1}=(-1)^{n+1}\braket{\lambda_n}{\psi_T}$ when the eigenvectors are ordered by decreasing eigenvalue.

As an aside, we mention that, in a similar fashion to perfect state transfer \cite{marais2013,godsil2012,jonckheere2015}, arbitrarily accurate solutions to the state synthesis problem are far more common. If we can find a chain for which $\braket{\lambda_n}{\psi_T}=\pm\lambda_{n,1}$ for all $n$, and the ratios of differences of eigenvalues are all irrational, then we can always wait long enough for the different phases to approximate the pattern $e^{-i\lambda_n t}=\braket{\lambda_n}{\psi_T}/\lambda_{n,1}$, and the analysis of the typical transfer time in \cite{jonckheere2015} is similarly applicable here. However, unlike perfect state transfer (where a symmetry condition arises naturally), it is not {\em a priori} clear how to fix the conditions $\braket{\lambda_n}{\psi_T}=\pm\lambda_{n,1}$. That is the main challenge that this work addresses. Our philosophy here, therefore, is to start from chains where we know this is true for some different target state ($\ket{N}$); the perfect state transfer chains, and to learn how to modify them appropriately for the true target state, while focussing on perfect solutions at a well-defined time rather than arbitrarily accurate solutions at an ill-defined time. Moreover, since the satisfying spectra for perfect state synthesis are discrete, we will select a fixed spectrum, and work constantly with that. We will rely extensively on the Lanczos algorithm, outlined briefly in the next subsection, to propagate any alterations that we make to the entire chain, ensuring that the spectrum of the system is kept fixed at this discrete choice.

\subsection{Lanczos Algorithm}

We will make use of the standard Lanczos Algorithm in our constructions \cite{gladwell2005}. This is an iterative algorithm which, at each step, takes as input the eigenvalues $\{\lambda_n\}$, the eigenvector elements at a particular site $m$, $\lambda_{n,m}$, and the coupling strength $J_{m-1}$ ($J_0=0$ to get the algorithm started). First, it calculates the magnetic field
$$
B_m=\bra{m}H\ket{m}=\sum_{n=1}^N\lambda_n\lambda_{n,m}^2,
$$
then uses that to give the next coupling strength, $J_m$:
$$
B_m^2+J_m^2+J_{m-1}^2=\bra{m}H^2\ket{m}=\sum_{n=1}^N\lambda_n^2\lambda_{n,m}^2.
$$
Finally, we use the eigenvector relations to derive the next eigenvector elements,
$$
\lambda_{n,m+1}=\frac{(\lambda_n-B_m)\lambda_{n,m}-J_{m-1}\lambda_{n,m-1}}{J_m}
$$
so that we have the required inputs for the next step of the algorithm. In this way, we can derive all the parameters of the Hamiltonian, and the eigenvectors, starting from a desired spectrum and the eigenvector amplitudes on the first site of the chain.

This construction has been used extensively in the study of perfect state transfer, with the connection first being realised in \cite{karbach2005}. Indeed, having established the necessary and sufficient conditions for perfect state transfer \cite{kay2010a}, all solutions are either found as analytic solutions, such as \cite{christandl2004,albanese2004}, or by fixing the spectrum and solving the Lanczos algorithm. The iteration is simply started by recognising that a perfect state transfer chain must have symmetric couplings, and so once a spectrum is fixed, that fixes the $\lambda_{n,1}$.

\section{Modifying Perfect State Transfer}\label{sec:analysis1}

For the task specified by Eq.\ \eqref{eqn:synthesis}, we have established that the eigenvalues of $H$ are tightly constrained -- it must be that $\braket{\lambda_n}{\psi_T}=\pm\lambda_{n,1}$ and $e^{-i\lambda_nt_0}=\pm e^{i\phi}$. We are going to select a particular spectrum that satisfies these conditions. Since they are reminiscent of the necessary and sufficient conditions for perfect state transfer \cite{kay2010a} (the ordered eigenvalues $\lambda_n>\lambda_{n+1}$ fulfil $e^{-i\lambda_n\pi/2}=(-1)^{n+1}$ with $t_0=\pi/2$), we proceed by assuming that $e^{-i\lambda_n\pi/2}=(-1)^{n+1}$. Under this assumption, every satisfying choice of $\{\lambda_n\}$ corresponds uniquely to a perfect state transfer Hamiltonian $\tilde H$, with fields $\tilde B_n$ and coupling strengths $\tilde J_n$.

There is no reason that one {\em has} to start by assuming the connection to a perfect state transfer system. Any existing solution that satisfies the eigenvalue conditions $e^{i\phi+i\lambda_nt_0}=\pm 1$ will do, at the cost of making the calculations slightly more complex. However, they will naturally lend themselves to different state synthesis tasks, specifically being able to produce outcomes that are in some sense close to the state produced by $\tilde H$. Since such states will typically be superpositions of the single excitation across many sites, which we already know how to address via different insights \cite{kay2017}, it makes most sense to concentrate on $\tilde H$ being a perfect state transfer Hamiltonian, and attempting to modify it in order to create superpositions of states on just a small number of sites.

{\it Example:} For the case $N=5$, we can select the spectrum to be $\{4,2,0,-2,-4\}$. There is a corresponding perfect state transfer Hamiltonian
$$
\tilde H=\left(\begin{array}{ccccc} 0 & 2 & 0 & 0 & 0 \\ 2 & 0 & \sqrt{6} & 0 & 0 \\ 0 & \sqrt{6} & 0 & \sqrt{6} & 0 \\ 0 & 0 & \sqrt{6} & 0 & 2 \\ 0 & 0 & 0 & 2 & 0 \end{array}\right).
$$

We will use these $\tilde H$ as the starting point for our solutions. They can be used to define a basis\footnote{To prove that {$\ket{\tilde v_m}$} forms a basis, write the elements out as columns of a matrix. If the matrix has non-zero determinant, the vectors span the space. Taking out a common non-zero factor {$\tilde\lambda_{n,1}$} from each row $n$ returns a matrix that is just the eigenvectors of $\tilde H$, which are all mutually orthogonal, and therefore has non-zero determinant.}
$$
\ket{\tilde v_m}=\sum_{n=1}^N\tilde\lambda_{n,1}\tilde\lambda_{n,m}\ket{n}
$$
for $m=1,\ldots,N$. Similarly, the state synthesis Hamiltonian $H$ has a basis
$$
\ket{v_n}=\sum_{k=1}^N\lambda_{k,1}\lambda_{k,n}\ket{k}.
$$
The choice of these bases is one of mathematical convenience, and does not exactly correspond to anything physical. That said, they clearly encapsulate the information about the two systems in a very useful way, facilitating the calculations of functions such as $\bra{1}f(H)\ket{n}$ simply by evaluating
$$
\left(\sum_{k=1}^Nf(\lambda_k)\bra{k}\right)\ket{v_n}.
$$
This includes normalisation ($f(H)=\identity$) and time evolution ($f(H)=e^{-iHt}$). Furthermore, one naive method for implementing the conditions that we want is to simultaneously solve
$$
\bra{1}H^k\ket{1}=\bra{\psi_T}H^k\ket{\psi_T}
$$
for $k=1,2,\ldots N$, which is closely connected. Indeed, our method essentially reduces to this calculation, except that our formalism will lend itself to finding instances in which the calculations are vastly easier to perform .

By definition, one basis can be written in terms of the other. We use the coefficients $\beta^{(n)}_m$,
$$
\ket{v_n}=\sum_{m=1}^N\beta^{(n)}_m\ket{\tilde v_m},
$$
which we often write as a table, $m$ specifying the rows, and $n$ the columns. Our aim is to find the vector $\ket{v_1}$. This contains the elements $\lambda_{n,1}^2$ which, together with the target spectrum, are the inputs for the Lanczos algorithm, and will thus specify $H$. In practice, this will be expressed by the $\beta^{(1)}_m$ and the (known) eigenvectors of $\tilde H$. 

Many of the coefficients $\beta^{(n)}_m$ can be predetermined. For instance, we can write that
$$
\left(\sum_{k=1}^N \bra{k}\right)\ket{v_m}=\sum_k\lambda_{1,k}\lambda_{m,k}.
$$
Since the eigenvectors are orthonormal, this satisfies
$$
\left(\sum_{k=1}^N \bra{k}\right)\ket{v_m}=\delta_{m,1},
$$
not only for the $\ket{v_m}$, but also the $\ket{\tilde v_m}$. But there is also the inter-conversion,
$$
\left(\sum_{k=1}^N \bra{k}\right)\ket{v_m}=\sum_{n=1}^N\beta^{(m)}_n\left(\sum_{k=1}^N \bra{k}\right)\ket{\tilde v_n},
$$
leaving us with
\begin{equation}
\delta_{m,1}=\beta^{(m)}_1.	\label{eq:toprow}
\end{equation}
Thus, the top row of the $\beta$-table is all zeros, except for the first element.

Similarly, we can compare the perfect state transfer conditions for $\tilde H$ to the state synthesis conditions of $H$. The state transfer condition may be written as
$$\bra{1}e^{-i\tilde H t_0}\ket{n}=\delta_{n,N}.$$ 
In terms of the eigenvectors, this is
\begin{equation}
\delta_{n,N}=\sum_k\lambda_{k,1}\lambda_{k,n}(-1)^{k+1}=\sum_k(-1)^{k+1}\braket{k}{\tilde v_n},	\label{eqn:subs}
\end{equation}
recalling that the evolution phase is alternately $\pm 1$.
Meanwhile,
$$
\alpha_n=\bra{1}e^{-iH t_0}\ket{n}
$$
can similarly be expressed as
$$
\alpha_n=\sum_k(-1)^{k+1}\braket{k}{v_n}.
$$
Again, we expand the two bases in terms of each other,
$$
\alpha_n=\sum_k(-1)^{k+1}\sum_m\beta^{(n)}_m\braket{k}{\tilde v_m}.
$$
Substituting Eq.\ \eqref{eqn:subs} yields
$$
\alpha_n=\sum_m\beta^{(n)}_m\delta_{m,N}=\beta^{(n)}_N.
$$
Thus, the bottom row of the $\beta$-table is simply the target amplitudes.
 
The entries of the $\beta$-table are related via
\begin{equation}\label{eqn:recursive}
(H\otimes\identity-\identity\otimes\tilde H)\sum_{n,m}\beta^{(n)}_m\ket{n,m}=0.
\end{equation}
A full derivation is given in the Appendix. This imposes a consistency condition for each element of the $\beta$-table. Applying it to the top-row condition of Eq.\ \eqref{eq:toprow} reveals that $\beta^{(n)}_m=0$ if $n>m$. Consequently, the right-hand column now reads $\ket{v_N}=\alpha_N\ket{\tilde v_N}$. Contiguous sets of 0s on the bottom row can also be propagated upwards using these relations, as demonstrated in the following example. Resolving all these consistency conditions yields all the system parameters of the solution.

A further necessary condition on the state synthesis task is $\alpha_N\neq 0$. This is a result of applying Eq.\ \eqref{eqn:recursive} for the element $\ket{n,n-1}$ ($n>1$), which implies
\begin{equation}
\beta^{(n)}_n\prod_{m=1}^{n-1}\tilde J_m=\prod_{m=1}^{n-1}J_m.	\label{eqn:Jrelation}
\end{equation}
For a chain of length $N$, we require $J_n\neq 0$ for all $n=1,\ldots,N-1$, discounting the possibility of producing two distinct chains. Thus, $\alpha_N=\beta^{(N)}_N\neq 0$; the synthesised state must have overlap with the end qubit. 

{\it Example:} For $N=5$, we aim to create an evolution $\ket{1}\rightarrow(\ket{4}+\ket{5})/\sqrt{2}$ in a time $t_0=\pi/2$ using the spectrum $\{4,2,0,-2,-4\}$. The $\beta$-table has the structure:
\begin{equation}
\begin{tabular}{c|ccccc}
$m\backslash n$ & 1 & 2 & 3 & 4 & 5 \\
\hline
1 & 1 & 0 & 0 & 0 & 0 \\
2 & $\beta^{(1)}_2$ & $\beta^{(2)}_2$ & 0 & 0 & 0 \\
3 & 0 & $\beta^{(2)}_3$ & $\beta^{(3)}_3$ & 0 & 0 \\
4 & 0 & 0 &  $\beta^{(3)}_4$ & $\beta^{(4)}_4$ & 0 \\
5 & 0 & 0 & 0 & $\frac{1}{\sqrt{2}}$ & $\frac{1}{\sqrt{2}}$
\end{tabular} \label{eqn:table}
\end{equation}
This is complete except for evaluation of the consistency conditions (Eq.\ \eqref{eqn:recursive}) on the four diagonals $\ket{n,n+k}$ for all $n$ and $k=-1,0,1,2$. As we'll see in Sec.\ \ref{sec:fractional}, it is not necessary to complete all these values, but for the sake of exposition, we evaluate the consistency conditions on the diagonals $k=-1,2$. These reveal that
$$
\beta^{(n)}_n=\prod_{m=1}^{n-1}\frac{J_m}{\tilde J_m}\qquad\text{and}\qquad \beta^{(n)}_{n+1}=\frac{1}{\sqrt{2}}\prod_{m=n}^{N-2}\frac{J_m}{\tilde J_{m+1}},
$$
with $\prod_{n=1}^{N-1}J_n=\prod_{n=1}^{N-1}\tilde J_n/\sqrt{2}=12\sqrt{2}$. The remaining consistency conditions, on the diagonals $k=0,1$ then yield
\begin{eqnarray*}
\beta^{(n)}_n&=&\frac{1}{\sqrt{2}}\prod_{m=n}^{N-2}\frac{J_m}{\tilde J_{m+1}}\frac{\sum_{m=n}^{N-1}B_m-2B_N}{\tilde J_n}	\\
\beta^{(n)}_{n+1}&=&\prod_{m=1}^{n-1}\frac{J_m}{\tilde J_m}\frac{\sum_{m=1}^nB_m}{\tilde J_n}.
\end{eqnarray*}
Simultaneous solution (eventually) fixes the relevant couplings to be
\begin{widetext}
\begin{equation}
H=\left(
\begin{array}{ccccc}
 -2\sqrt{\frac{1}{13} \left(6-\sqrt{10}\right)} & -2\sqrt{\frac{1}{13} \left(7+\sqrt{10}\right)} & 0 & 0 & 0
   \\
 -2\sqrt{\frac{1}{13} \left(7+\sqrt{10}\right)} & -\sqrt{\frac{5}{13} \left(62-19
   \sqrt{10}\right)} & \sqrt{9 \sqrt{10}-24} & 0 & 0 \\
 0 & \sqrt{9 \sqrt{10}-24} & -\sqrt{\frac{5}{13} \left(118-37 \sqrt{10}\right)} &
   2\sqrt{\frac{2}{13} \left(1+2 \sqrt{10}\right)} & 0 \\
 0 & 0 &2 \sqrt{\frac{2}{13} \left(1+2 \sqrt{10}\right)} &\sqrt{\frac{1}{26} \left(62-19
   \sqrt{10}\right)} & -\sqrt{3+\sqrt{\frac{5}{2}}} \\
 0 & 0 & 0 & -\sqrt{3+\sqrt{\frac{5}{2}}} & \sqrt{3+\sqrt{\frac{5}{2}}} \\
\end{array}
\right).\label{eqn:example}
\end{equation}
\end{widetext}

\section{Fractional Revivals} \label{sec:fractional}

Generically, the values $\{\beta^{(1)}_m\}$ are hard to derive in terms of the $\alpha_n$. However, the purpose of selecting the basis $\ket{\tilde v_m}$ for decomposing $\ket{v_1}$ is that certain special cases of particular interest are not as hard as the generic case. We now specialise to the evolution
$$
\ket{1}\xrightarrow{e^{-iH\pi/2}}\alpha_1\ket{1}+\alpha_r\ket{r}+\alpha_N\ket{N}.
$$
for $r\neq 1,N$. In this case, since $\lambda_{n,1}(-1)^{n+1}=\braket{\psi_T}{\lambda_n}$, we can multiply by $\lambda_{n,1}$ and use that $\ket{v_N}=\alpha_N\ket{\tilde v_N}$:
$$
 \left(\sum_n(-1)^{n+1}\proj{n}\right)\ket{v_1}=\alpha_1\ket{v_1}+\alpha_r\ket{v_r}+\alpha_N^2\ket{\tilde v_N}.
 $$
To evaluate this, note that
\begin{eqnarray*}
\left(\sum_n(-1)^{n+1}\proj{n}\right)\ket{\tilde v_m}&=&\sum_n\lambda_{1,n}\lambda_{m,n}(-1)^{n+1}	\\
&=&\ket{\tilde v_{N+1-m}},
\end{eqnarray*}
using the symmetry property of eigenvectors in perfect state transfer chains. Thus,
$$
\alpha_1\ket{v_1}+\alpha_r\ket{v_r}+\alpha_N^2\ket{\tilde v_N}=\sum_n\beta^{(1)}_m\ket{\tilde v_{N+1-m}}.
$$
The basis on the $\ket{\tilde v_n}$ can safely be relabelled to make it easier to work with:
\begin{equation}
\alpha_r\sum_{m=1}^N\!\beta^{(r)}_m\ket{m}=(S-\alpha_1\identity)\sum_{m=1}^N\!\beta^{(1)}_m\ket{m}-\alpha_N^2\ket{N}, \label{eq:small}
\end{equation}
where $S=\sum_{n=1}^N\ket{n}\bra{N+1-n}$. This relationship will permit us to derive the desired $\beta^{(1)}_m$.

With $\alpha_r=0$, one quickly recovers the standard instances of perfect revivals \cite{kay2010a} -- it requires $\beta^{(1)}_{N+1-m}=\alpha_1\beta^{(1)}_{m}$ if $m\neq N$, and hence $\beta^{(1)}_m=0$ for $m=2,\ldots,N-1$, $\beta^{(1)}_1=1$ and $\beta^{(1)}_N=\alpha_1$. Thus, we have that
$$
\ket{v_1}=\ket{\tilde v_1}+\alpha_1\ket{\tilde v_N},
$$
allowing us to identify that
$$
\lambda_{n,1}^2=\tilde\lambda_{n,1}(\tilde\lambda_{n,1}+\alpha_1\tilde\lambda_{n,N}).
$$
For the standard solution of spin chains \cite{christandl2004}, we have the analytic expression for the $\tilde\lambda_{n,1}$ of
$$
\tilde\lambda_{n,1}=\tilde\lambda_{n,N}(-1)^{n+1}=\frac1{2^{(N-1)/2}}\sqrt{\binom{N-1}{n-1}}.
$$
From here, the Lanczos algorithm proceeds as normal.

Eq.\ \eqref{eq:small} is particularly compelling when $r$ is large. For $r=N-1$, we recall that most of the column of the $\beta$-table has already been completed:
$$
\sum_{m=1}^N\!\beta^{(N-1)}_m\ket{m}=\alpha_{N-1}\ket{N}+\beta^{(N-1)}_{N-1}\ket{N-1},
$$
i.e.\ there is only one undetermined value $\beta^{(N-1)}_{N-1}$. Upon analysing
\begin{multline*}
(\alpha_N^2+\alpha_{N-1}^2)\ket{N}+\alpha_{N-1}\beta^{(N-1)}_{N-1}\ket{N-1}\\=(S-\alpha_1\identity)\sum_m\beta^{(1)}_m\ket{m}
\end{multline*}
for $m=1,2,\ldots N-2$, we have $\alpha_1\beta^{(1)}_m=\beta^{(1)}_{N+1-m}$ and, indeed, $\beta^{(1)}_n=0$ for $n=3,4,\ldots ,N-2$. Since $\beta^{(1)}_1=1$, the only undetermined parameter is $\beta^{(1)}_2$: 
$$
\ket{v_1}=\ket{\tilde v_1}+\alpha_1\ket{\tilde v_N}+\beta^{(1)}_2(\ket{\tilde v_2}+\alpha_1\ket{\tilde v_{N-2}}).
$$
Furthermore, this parameter can be straightforwardly evaluated using normalisation -- since $\ket{v_N}=\alpha_N\ket{\tilde v_N}$, it follows that
$$
1=\sum_n\braket{N}{\lambda_n}^2=\sum_n\frac{\alpha_N^2\braket{n}{\tilde v_N}^2}{\braket{n}{v_1}}.
$$
Substituting the definitions reveals that
$$
1=\alpha_N^2\sum_{n=1}^N\frac{\tilde\lambda_{n,1}^2}{(1+\alpha_1(-1)^{n+1})\left(1+\beta^{(1)}_2\frac{\lambda_n-\tilde B_1}{\tilde J_1}\right)},
$$
where we have invoked the symmetry property for perfect state transfer of $\lambda_{n,m}=(-1)^{n+1}\lambda_{n,N+1-m}$ and the relation $\tilde J_1\tilde\lambda_{n,2}+\tilde B_1\tilde\lambda_{n,1}=\lambda_n\tilde\lambda_{n,1}$. This is equivalent to solving
$$
\bra{1}\frac{1}{(1-\tilde B_1\beta^{(1)}_2/\tilde J_1)\identity+\beta^{(1)}_2\tilde H/\tilde J_1}(\ket{1}-\alpha_1\ket{N})=\frac{1-\alpha_1^2}{\alpha_N^2},
$$
which always has a real solution. 

\begin{figure}
\begin{center}
\includegraphics[width=0.45\textwidth]{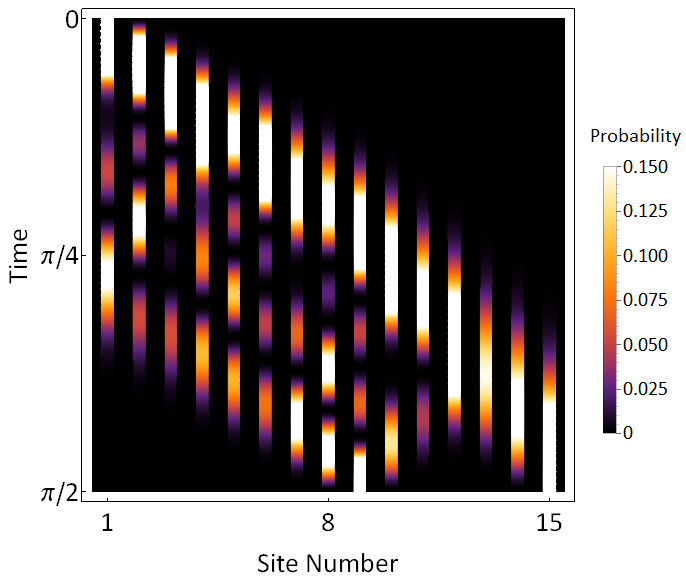}
\end{center}
\vspace{-0.5cm}
\caption{A 15 qubit chain undergoes the evolution $\ket{1}\rightarrow(\ket{9}+\ket{15})/\sqrt{2}$. The parameters in the first column $\beta^{(1)}_m$ with $m=3,5,7$ are initially unknown.}\label{fig:fractional}
\vspace{-0.5cm}
\end{figure}

{\it Example:} Returning to the previous example ($\alpha_4=\alpha_5=1/\sqrt{2}$), we take the $\beta$-table of Eq.\ \eqref{eqn:table} and acknowledge that $\lambda_{n,1}^2=\tilde\lambda_{n,1}^2+\beta^{(1)}_2\tilde\lambda_{n,1}\tilde\lambda_{n,2}$ with $\tilde\lambda_{n,1}^2=\frac{1}{16}\binom{4}{n-1}$ and $\tilde\lambda_{n,2}=\tilde\lambda_{n,1}(3-n)$. Since $\ket{v_5}=\alpha_5\ket{\tilde v_5}$,
$$
\lambda_{n,5}^2=\frac{1}{2}\frac{\tilde\lambda_{n,1}^2}{1+\beta^{(1)}_2(3-n)},
$$
to which we apply the normalisation condition
$$
32=\sum_{n=0}^4\frac{\binom{4}{n}}{1+\beta^{(1)}_2(2-n)}.
$$
This simplifies to $52{\beta^{(1)}_2}^4-48{\beta^{(1)}_2}^2+8=0,$ i.e.\ ${\beta^{(1)}_2}^2=(6\pm\sqrt{10})/13$. Having found $\beta^{(1)}_2$, and consequently the $\lambda_{n,1}$, the Lanczos algorithm can be applied, starting with
\begin{eqnarray*}
\sum_{n=1}^5\lambda_n\lambda_{n,1}^2&=&B_1=\tilde B_1+\tilde J_1\beta^{(1)}_2	\\
\sum_{n=1}^5\lambda_n^2\lambda_{n,1}^2&=&B_1^2+J_1^2=\tilde B_1^2+\tilde J_1^2+\beta^{(1)}_2\tilde J_1(\tilde B_1+\tilde B_2).
\end{eqnarray*}
The whole procedure iterates to calculate all the values shown in Eq.\ \eqref{eqn:example}.

As $r$ decreases, the number of parameters increases correspondingly, rendering the solution more difficult to derive. However, if $\alpha_{N+1-2m}=0$ for all $m$, then the complexity can be reduced by assuming that $\beta^{(n)}_m=0$ for all $n+m$ even (which also imposes that $B_n=\tilde B_n$ for all sites). Fig.\ \ref{fig:fractional} depicts the evolution of a 15 qubit system designed to implement $\ket{1}\xrightarrow{e^{-i\pi H/2}}(\ket{9}+\ket{15})/\sqrt{2}$. 

\begin{figure}
\begin{center}
\includegraphics[width=0.45\textwidth]{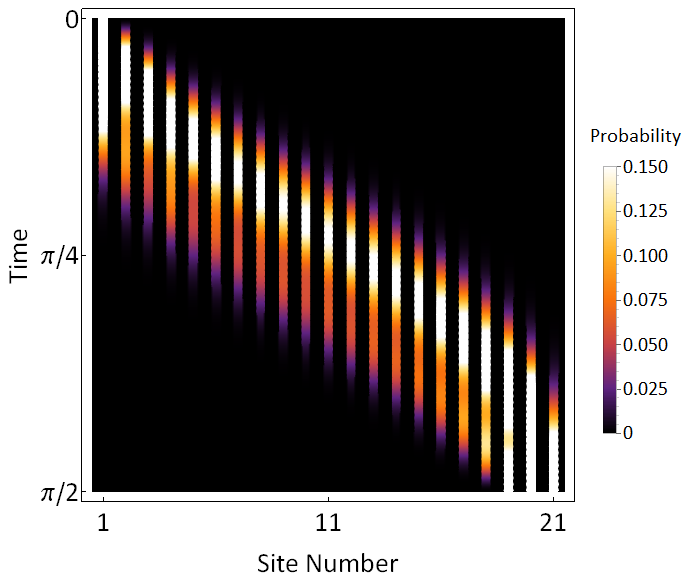}
\end{center}
\vspace{-0.5cm}
\caption{Evolution $\ket{1}\rightarrow(\ket{19}+\ket{20}+\ket{21})/\sqrt{3}$.}\label{fig:triple}
\vspace{-0.5cm}
\end{figure}

More generally, if the last $k$ amplitudes (and $\alpha_1$) are to be non-zero, then the first $k$ coefficients ($k<N/2$) $\beta^{(1)}_k$ are non-zero in $\ket{v_1}$ and $\beta^{(1)}_{N+1-k}=\alpha_1\beta^{(1)}_k$ for the last $k$ coefficients, with the rest 0. For sufficiently small $k$, we can solve for these through normalisation considerations, and relations such as $\alpha_{N-1}J_{N-1}=-\alpha_NB_N$ (Eq.\ \eqref{eqn:recursive} applied to the element $\ket{N,N}$), calculating $J_{N-1}$ and $B_N$ from $\ket{v_N}$. An example is depicted in Fig.\ \ref{fig:triple} that superposes the initial excitation equally over the last 3 sites of the chain.

At the extreme of small $r$, we relate the $\beta^{(1)}_m$ and $\beta^{(r)}_m$ via Eq.\ \eqref{eqn:recursive}. For instance,
$$
\sum_{m=1}^N\beta^{(2)}_m\ket{m}=\frac{\tilde H-B_1\identity}{J_1}\sum_{m=1}^N\beta^{(1)}_m\ket{m},
$$
revealing a linear system for the $\beta^{(1)}_m$ parametrised by only $B_1$ and $J_1$. Enforcing $\beta^{(1)}_1\pm\beta^{(1)}_N=1\pm\alpha_1$ determines these values. The case of $r=3$ constitutes the starting point for the example given in Fig.\ \ref{fig:mid}. The different techniques for either end of the chain can be combined to create transfers such as $\ket{1}\rightarrow(\ket{1}+\ket{2}+\ket{N-1}+\ket{N})/2$.

\section{Transfer from Middle} \label{sec:middle} 
\begin{figure}
\begin{center}
\includegraphics[width=0.45\textwidth]{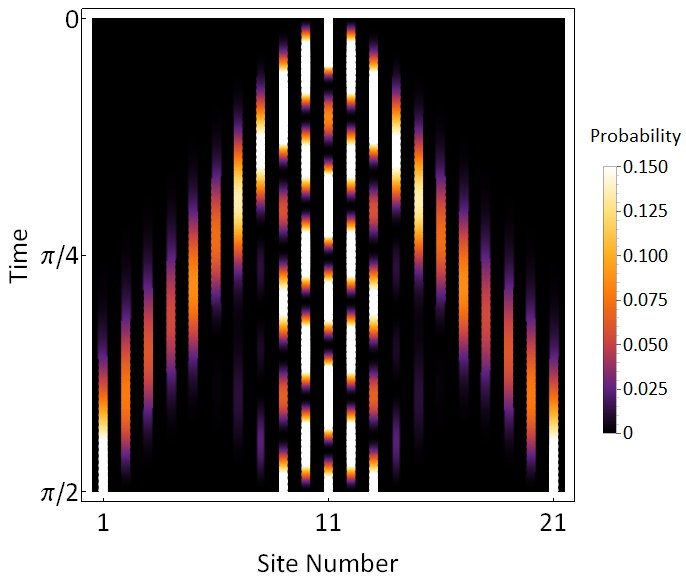}
\end{center}
\vspace{-0.5cm}
\caption{Evolution $\ket{11}\rightarrow(\ket{1}+\ket{9}+\ket{11}+\ket{13}+\ket{21})/\sqrt{5}$.}\label{fig:mid}
\vspace{-0.5cm}
\end{figure}

Our constructions so far are good at creating perfect revivals that are localised at the ends of the chain, but not in the middle. However, we can make use of an observation that originates in \cite{plenio2004,kay2005} to modify the $N\times N$ matrix $H$ which creates the evolution $\ket{1}\xrightarrow{e^{-iHt_0}}\ket{\psi_T}$. We construct a new Hamiltonian, $H'$ of $2N-1$ qubits, satisfying $B_n'=B_{1+|n-N|}$ and
$$
J_n'=\left\{\begin{array}{cc}J_{\half+\left|n-N+\half\right|}	&n\neq N-1,N	\\
J_1\cos\theta & n=N-1	\\
J_1\sin\theta & n=N
\end{array}\right..
$$
This generates the evolution $\ket{N}\xrightarrow{e^{-iH't_0}}\sum_{n=1}^{N}\alpha_{n}\ket{n'}$, where $\ket{n'}=\cos\theta\ket{N+1-n}+\sin\theta\ket{N-1+n}$ for $n=1,\ldots,N-1$ and $\ket{1'}=\ket{N}$, thereby facilitating production of a superposition over the ends and the middle of the chain. One can readily see that the subspace of $H'$ spanned by $\{\ket{n'}\}$ is exactly the matrix $H$. For example, in Fig.\ \ref{fig:mid}, we constructed a chain of length 11 with the evolution $\ket{1}\rightarrow(\ket{1}+\sqrt{2}\ket{3}+\sqrt{2}\ket{11})/\sqrt{5}$\footnote{There are two qualitatively different solutions.}, and produced a corresponding $H'$ of 21 sites that achieves $\ket{11}\rightarrow(\ket{1}+\ket{9}+\ket{11}+\ket{13}+\ket{21})/\sqrt{5}$.  The further advantage is in speed; it will typically take about half the time to generate a particular state starting from the middle rather than one end because the excitation only has half as far to go.

\section{Numerical Approach}\label{sec:numerics} With a limited range of analytic solutions, we seek numerical techniques for generating a wider range of evolutions. A perturbative scheme for the $\{\beta^{(1)}_m\}$, as opposed to examining the Hamiltonian perturbation, has the advantage of being isospectral by construction, with correspondingly fewer parameters to determine. A first order perturbative expansion is easily applied to Eq.\ \eqref{eqn:recursive} provided one knows how the $J_n$ and $B_n$ are perturbed. These shifts may be derived from the identities
\begin{eqnarray*}
B_n&=&\frac{\bra{1}\tilde H^n\ket{\beta^{(n)}}}{\bra{1}\tilde H^{n-1}\ket{\beta^{(n)}}}-\frac{\bra{1}\tilde H^{n-1}\ket{\beta^{(n-1)}}}{\bra{1}\tilde H^{n-2}\ket{\beta^{(n-1)}}}	\\
J_n^2&=&\frac{\bra{1}\tilde H^{n+1}\ket{\beta^{(n)}}}{\bra{1}\tilde H^{n-1}\ket{\beta^{(n)}}}-\frac{\bra{1}\tilde H^n\ket{\beta^{(n-1)}}}{\bra{1}\tilde H^{n-2}\ket{\beta^{(n-1)}}}	\\
&&-B_n\frac{\bra{1}\tilde H^n\ket{\beta^{(n)}}}{\bra{1}\tilde H^{n-1}\ket{\beta^{(n)}}}
\end{eqnarray*}

\begin{figure}
\begin{center}
\includegraphics[width=0.45\textwidth]{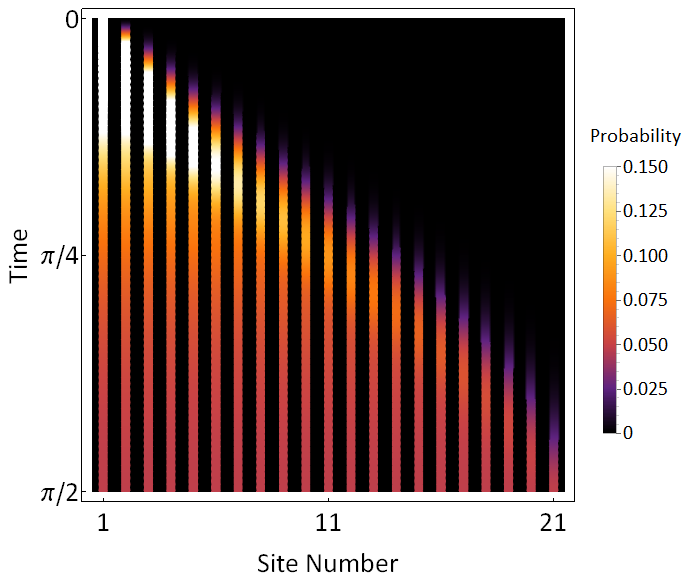}
\end{center}
\vspace{-0.5cm}
\caption{An excitation initially localised on site one of a 21 spin chain evolves into a $W$ state over all sites.}\label{fig:Wstate1}
\end{figure}

\noindent where $\ket{\beta^{(n)}}=\sum_m\beta^{(n)}_m\ket{m}$. Practically, this involves ensuring that $\delta\beta^{(n)}_{n-1}=\delta\beta^{(n)}_{n-2}=0$ for all $n$. To tolerate the high degree of non-linearity in the system, a good initial guess is essential. The choice
$$
\braket{n}{v_1}=\frac{r^n(1-r)}{1-r^N} \text{ where }\frac{(1-r)(1+r^N)}{(1+r)(1-r^N)}=\alpha_1,
$$
uniform spectrum $\lambda_n=(N+1)-2n$, and $t_0=\pi/2$, yields an output state with a roughly uniform spread of amplitudes for chains of length up to $N\approx 50$, while fixing $\beta^{(1)}_N=\alpha_1$. When $N=21$, this choice produces an output $\ket{\psi_{\text{out}}}$ with $\braket{\psi_{\text{out}}}{\psi_{T}}=0.985$ where $\ket{\psi_T}$ is the $W$-state. This is close enough that a perturbative approach stands a good chance of converging. Fig.\ \ref{fig:Wstate1} depicts the evolution of one such system, whose value $\braket{\psi_{\text{out}}}{\psi_{T}}>1-10^{-24}$.

\begin{figure}[!btp]
\begin{center}
\includegraphics[width=0.45\textwidth]{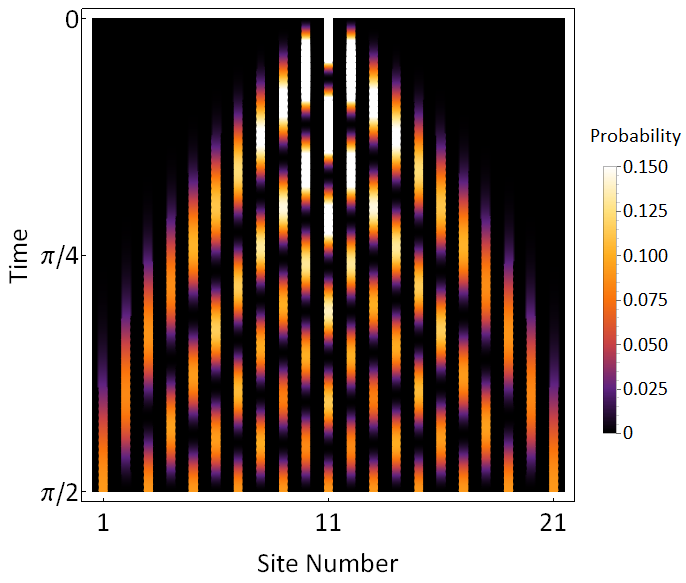}
\end{center}
\vspace{-0.5cm}
\caption{Evolution of $\ket{11}\rightarrow \sum_{n=1}^{11}\ket{2n-1}/\sqrt{11}$ on a chain of length 21, designed by the perturbative methods.}\label{fig:WstateMidOdd}
\vspace{-0.5cm}
\end{figure}

As before, if the target state has $\alpha_{N+1-2n}=0$ for all $n$, one can assume that $\beta^{(n)}_m=0$ for all $n+m$ odd, imposing that $B_n=0$, and reducing the number of parameters. To generate Fig.\ \ref{fig:WstateMidOdd}, we produced a chain that achieved $\ket{1}\rightarrow(\ket{1}+\sqrt{2}\sum_{n=1}^{5}\ket{2n+1})/\sqrt{11}$ using these perturbative methods, and then modified it by our observation of Sec.\ \ref{sec:middle} to create the evolution $\ket{11}\rightarrow \sum_{n=1}^{11}\ket{2n-1}/\sqrt{11}$ on a system of size $N=21$.

\section{Speed of State Synthesis}\label{sec:speed}

\begin{figure}[!tb]
\begin{center}
\includegraphics[width=0.45\textwidth]{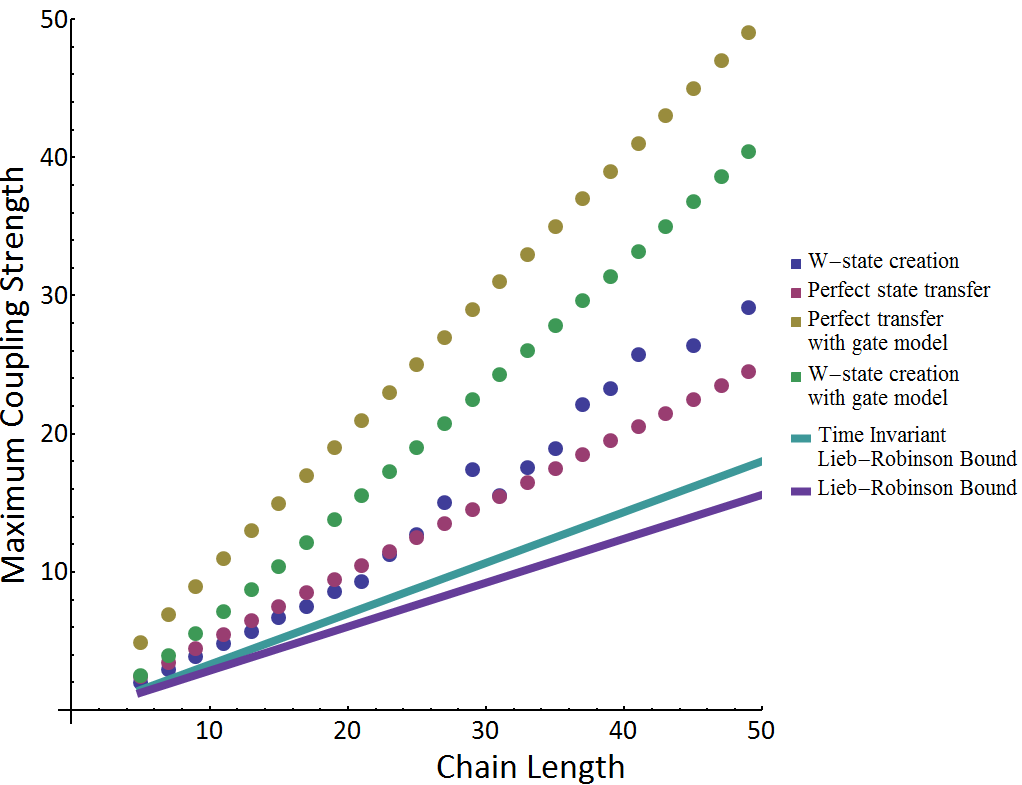}
\end{center}
\vspace{-0.5cm}
\caption{Results of numerical solutions to $W$-state generation and perfect state transfer in a fixed state synthesis time, $t_0=\pi/2$. Note that perfect state transfer is optimal \cite{yung2006} indicating the limitation of the bounds. These are compared to Lieb-Robinson bounds for the system, and the comparable results if a sequential set of quantum gates are applied. Smaller is better.}\label{fig:speed}
\vspace{-0.5cm}
\end{figure}

As is the case for perfect state transfer \cite{yung2006}, state synthesis is usually substantially quicker than via a gate decomposition that has the same locality constraints. For example, if $J_{\max}=\max\{J_n\}$, then the $W$-state synthesis example of Fig.\ \ref{fig:Wstate1} has $J_{\max}t_0=14.6$, while a sequence of consecutive swaps of strength $J_{\max}$ creating the transformations
$$
\sqrt{\frac{N+1-k}{N}}\ket{k}\rightarrow\frac{1}{\sqrt{N}}\ket{k}+\sqrt{\frac{N-k}{N}}\ket{k+1}
$$
has $J_{\max}t_0=23.0$. For other systems sizes, the values are given for comparison in Fig.\ \ref{fig:speed}.

We would now like to justify that our choice of spectrum leads to a near-optimal state synthesis time. Consider any spectrum that is compatible with state synthesis, i.e.\ $\lambda_n=\pi m_n/t_0$ where $m_n$ are distinct integers. Without loss of generality, one value is $m_k=0$ (simply shifting all eigenvalues by the same amount only changes the Hamiltonian by an irrelevant identity matrix). We can use the Lanczos algorithm to construct a symmetric $\tilde H$ (meaning that $\tilde B_{N+1-n}=\tilde B_n$ and $\tilde J_n=\tilde J_{N-n}$) with that spectrum and positive $\tilde J_n$. The $k^{th}$ eigenvector, corresponding to the zero eigenvalue, has weight $w=\braket{1}{\lambda_k}^2$ on the first site of the chain. From \cite{gladwell2005}, the coupling strengths are related to the eigenvalues via
$$
\prod_{n=1}^{N-1}\tilde J_n=w|p'(0)|,
$$
where $p(\lambda)$ is the characteristic polynomial. Hence
$$
\prod_{n=1}^{N-1}\tilde J_n=w\left(\frac{\pi}{t_0}\right)^{N-1}\left|\prod_{n\neq k}m_n\right|.
$$
The evolution of this Hamiltonian can be expressed as
$$
\ket{1}\xrightarrow{e^{-i\tilde Ht_0}}\sum_n\tilde\alpha_n\ket{n}.
$$
By virtue of a parallel derivation to Eq.\ \eqref{eqn:Jrelation}, a construction of $H$ from $\tilde H$ yields
$$
\prod_{n=1}^{N-1}J_n=\frac{\alpha_N}{\tilde\alpha_N}\prod_{n=1}^{N-1}\tilde J_n.
$$
This yields a simple inequality
$$
J_{\max}>\left(\prod_{n=1}^{N-1}J_n\right)^{1/(N-1)}=\frac{\pi}{t_0}\left(w\frac{\alpha_N}{\tilde\alpha_N}\left|\prod_{n\neq k}m_n\right|\right)^{1/(N-1)}.
$$
The smallest possible product of integers is $((N-1)/2)!^2$, and $\tilde \alpha_N$ is no larger than 1 (both corresponding to our chosen perfect state transfer chain), meaning
$$
J_{\max} t_0>\pi\left(w\alpha_N\left(\frac{N-1}{2}\right)\!\text{\Large!}^2\right)^{1/(N-1)}.
$$
In the large $N$ limit, Stirling's formula reveals that
$$
J_{\max} t_0>\frac{\pi}{2e}(N-1),
$$
independent of the target state, provided $w\alpha_N$ is not exponentially small\footnote{Of course, we select $\alpha_N$ to be a particular value, say $1/\sqrt{N}$ for the W-state.  For our chosen spectrum, $w=2^{1-N}\binom{N-1}{\frac{N-1}{2}}$, and is therefore not exponentially small.}. This is essentially a Lieb-Robinson bound for the system \cite{bravyi2006}, but is tighter than the general bounds, which numerically appear to give $J_{\max}t_0\geq (N-1)/2$ \cite{murphy2010}, by virtue of specialising to the time invariant case and specific form of the Hamiltonian. Nevertheless, the difference is astoundingly slim -- if solutions can be tight to the bound, there is little speed to be gained in moving from a fixed local Hamiltonian to one with arbitrary local controls! Without a useful bound on the value of $w$, the bounds only apply to the large size limit and cannot be adapted to the finite size case. Instead, we compare these Lieb-Robinson style bounds to the maximum coupling strength involved in two systems -- one that generates $W$ states, and one that performs perfect state transfer \cite{christandl2004}. These are depicted in Fig.\ \ref{fig:speed}.


\section{Robustness}\label{sec:robustness}

\begin{figure}[!tb]
\begin{center}
\includegraphics[width=0.45\textwidth]{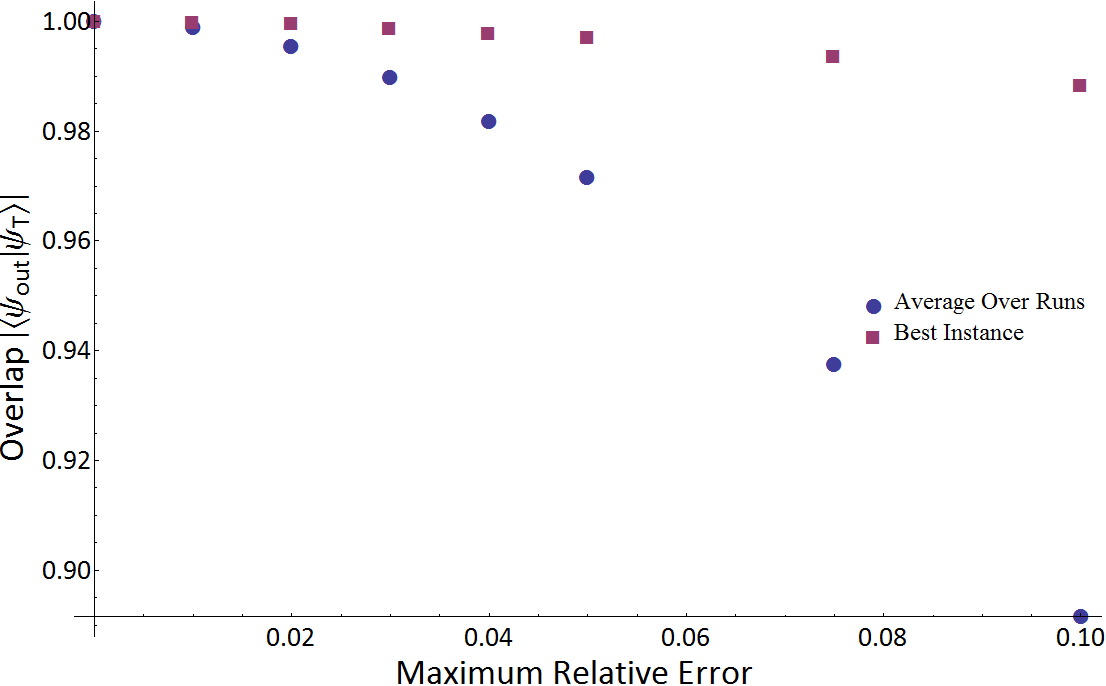}
\end{center}
\vspace{-0.5cm}
\caption{When the solution depicted in Fig.\ \ref{fig:Wstate1} for creating a W-state is perturbed, the output remains at high fidelity.}\label{fig:robustW}
\end{figure}

\begin{figure}[!tp]
\begin{center}
\includegraphics[width=0.45\textwidth]{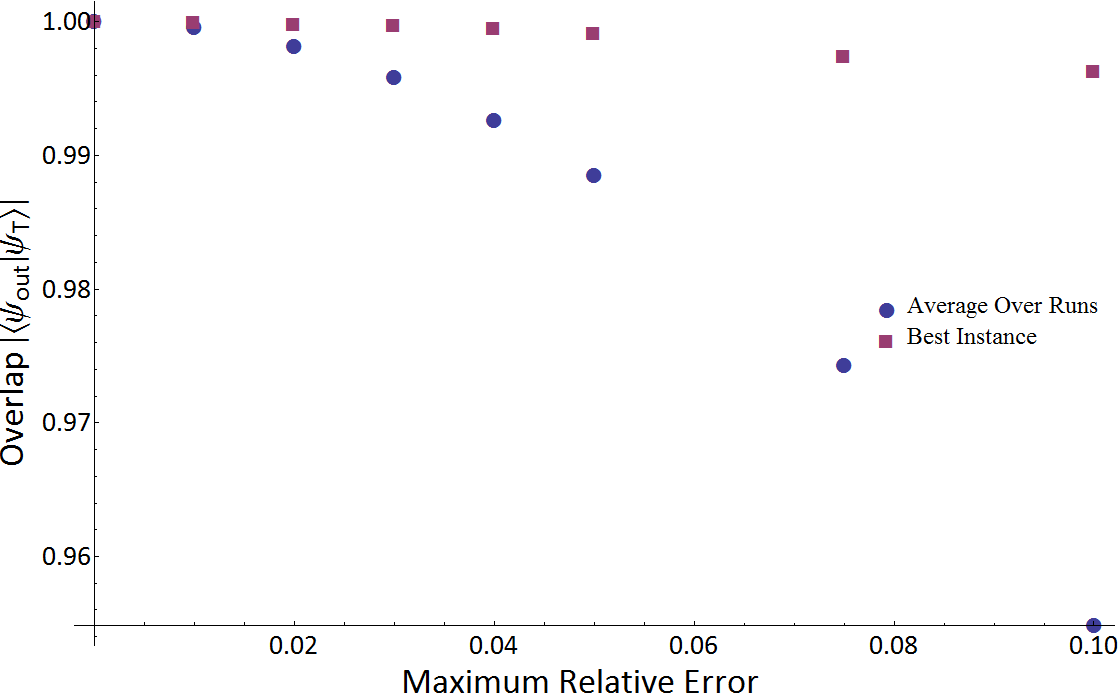}
\end{center}
\vspace{-0.5cm}
\caption{When the solution depicted in Fig.\ \ref{fig:fractional} for superposing sites 8 and 15 of a 15 qubit chain is perturbed, the output remains at high fidelity.}\label{fig:robustrevival}
\vspace{-0.5cm}
\end{figure}

Inevitably, any real experiment is imperfect, from inaccuracies in the intended coupling strengths and magnetic fields through to dynamic errors. In this section, we do not address the full spectrum of possibilities, merely aim to justify that the solutions presented so far have a basic level of robustness. To that end, we concentrate on manufacturing imperfections, shifting each coupling and magnetic field by a random small fraction. We compare the average arrival fidelity of the target state to the best out of 10000 realisations selected uniformly at random. While the average is what we might expect from the performance of any single instance, the advantage of prior manufacture of a fixed device is the facility to make several, test them, and choose the best. We examine two different, representative, cases. The first is the W-state production of Fig.\ \ref{fig:Wstate1}, depicted in Fig.\ \ref{fig:robustW}. The second is an analytic revival on two sites, chosen because, from the evolution depicted in Fig.\ \ref{fig:fractional}, one might anticipate a particular dependence upon intricate interferences, and therefore exhibit notable susceptibility to imperfections. Such concerns appear to be unfounded, see Fig.\ \ref{fig:robustrevival}.

\section{Conclusions} Many different cases of fractional revivals can be re-engineered from a perfect state transfer chain, meaning that a single excitation can be input at one end of a chain, and the natural dynamics evolve it into the desired superposition of that single excitation across a small number of sites, usually localised at either end of the chain. We have also described a perturbative technique that admits the possibility of moving beyond the analytically tractable cases and yet still produces useful coupling schemes for a variety of quantum state synthesis tasks. The solutions are robust against imperfections, and are near-optimal in speed for small system sizes. An important assumption is that all the amplitudes in the target state are real. Supporting calculations are provided via a Mathematica workbook \cite{kay2016d}.

Experimental prospects for this work are good. The basic technology of evanescently-coupled waveguides has already been applied to perfect state transfer \cite{Perez-Leija2013}. Moreover, the tasks considered here only involve a single excitation, not a superposition of states, so one does not require the additional lengths of more recent experiments \cite{Weimann2014,Chapman2016}. However, the efficacy of such a scheme would have to be compared to other methods such as \cite{grafe2014}.

We anticipate that a wide variety of other systems, with varying degrees of control, should also be capable of state synthesis, and exploring these is likely to be most beneficial to experiments. Another extremal case is a network of uniformly coupled spins. What network topologies permit the creation of states such as the $W$ state (aside from the trivial star network)? The basic properties, such as the necessary conditions, derived here will also be relevant \cite{kay2011a}.

{\em Acknowledgements:} We would like to thank L.\ Banchi and G.\ Coutinho for introductory conversations. This work was supported by EPSRC grant EP/N035097/1.

\onecolumn\newpage
\appendix

\section{Proof of Equation \eqref{eqn:recursive}}

Recall that for every eigenvalue $\lambda_n$, and every site $m$, the eigenvectors obey the conditions
$$
(\lambda_n-B_m)\lambda_{n,m}=J_{m-1}\lambda_{n,m-1}+J_m\lambda_{n,m}.
$$
If we multiply by $\lambda_{n,1}$, then this can be written as
\begin{equation}\label{eqn:app1}
(\lambda_n-B_m)\braket{n}{v_m}=J_{m-1}\braket{n}{v_{m-1}}+J_m\braket{n}{v_{m+1}}.
\end{equation}
We replace these in terms of the $\braket{n}{\tilde v_k}$, so
$$
\sum_k(\lambda_n-B_m)\beta^{(m)}_k\braket{n}{\tilde v_k}-J_{m-1}\beta^{(m-1)}_k\braket{n}{\tilde v_k}-J_m\beta^{(m+1)}_k\braket{n}{\tilde v_k}=0.
$$
Since Eq.\ \eqref{eqn:app1} holds for the system $H$ and $\tilde H$, so we also have
\begin{equation}\label{eqn:app2}
(\lambda_n-\tilde B_m)\braket{n}{\tilde v_m}=\tilde J_{m-1}\braket{n}{\tilde v_{m-1}}+\tilde J_m\braket{n}{\tilde v_{m+1}},
\end{equation}
and this can be used to eliminate the $\lambda_n\braket{n}{\tilde v_k}$ term:
$$
\sum_k\left((\tilde B_k-B_m)\beta^{(m)}_k-J_{m-1}\beta^{(m-1)}_k-J_m\beta^{(m+1)}_k+\beta^{(m)}_{k-1}\tilde J_{k-1}+\beta^{(m)}_{k+1}\tilde J_k\right) \braket{n}{\tilde v_k}=0.
$$
As this is true for all $n$, it simplifies to
$$
\sum_k\left((\tilde B_k-B_m)\beta^{(m)}_k-J_{m-1}\beta^{(m-1)}_k-J_m\beta^{(m+1)}_k+\beta^{(m)}_{k-1}\tilde J_{k-1}+\beta^{(m)}_{k+1}\tilde J_k\right)\ket{\tilde v_k}=0.
$$
Since the $\ket{\tilde v_k}$ form a basis, there is no linear combination that gives 0, so the only solution is that every coefficient, for every $k$ and $m$, is 0. This can succinctly be written as
$$
(H\otimes\identity-\identity\otimes\tilde H)\sum_{m,k}\beta^{(m)}_k\ket{m,k}=0,
$$
as required.
\end{document}